\documentstyle[preprint,aps]{revtex}
\begin{document}
%\twocolumn[\hsize\textwidth\columnwidth\hsize\csname@twocolumnfalse\endcsname

\title{ The Lightest Scalar Nonet as Higgs Bosons of Strong Interactions }
\author{Nils A. T\"ornqvist\footnote{\tt{e-mail:
nils.tornqvist@helsinki.fi}}}
\address{Department of Physical Sciences\\
University of Helsinki \\
POB 64, FIN--00014,  Finland}
%\date{\today}
\maketitle
\begin{abstract}
I discuss how an extra light scalar meson multiplet  could be
understood as an effective Higgs nonet of a hidden local $U(3)$
symmetry. There is growing evidence   that low energy data
requires  in addition to  a conventional $^3P_0$ $\bar q q$ nonet
near 1.4 GeV, another light scalar nonet-like structure below 1
GeV, ($\sigma(600)$, $a_0(980)$, $f_0(980)$, $\kappa$), which
could be interpreted as such a Higgs nonet.
 \\
\noindent Pacs numbers: 11.15.Ex, 11.30.Hv, 12.39.Fe, 14.80.Cp
\vskip 0.20cm
\end{abstract}

%]
\def \gam {\frac{ N_f N_cg^2_{\pi q\bar q}}{8\pi} }
\def \gam {\frac{ N_f N_cg^2_{\pi q\bar q}}{8\pi} }
\def \gamm {N_f N_cg^2_{\pi q\bar q}/(8\pi) }
\def \be {\begin{equation}}
\def \ba {\begin{eqnarray}}
\def \ee {\end{equation}}
\def \ea {\end{eqnarray}}
\def \gap {{\rm gap}}
\def \gapp {{\rm \overline{gap}}}
\def \gappp {{\rm \overline{\overline{gap}}}}
\def \im {{\rm Im}}
\def \re {{\rm Re}}
\def \Tr {{\rm Tr}}
\def \P {$0^{-+}$}
\def \S {$0^{++}$}
\def \uu {$u\bar u$}
\def \dd {$d\bar d$}
\def \ss {$s\bar s$}
\def \qq {$q\bar q$}
\def \qqq {$qqq$}
\def \lsm {L$\sigma$M}
\def \sig {$\sigma$}
\def \gam {\frac{ N_f N_cg^2_{\pi q\bar q}}{8\pi} }
\def \gamm {N_f N_cg^2_{\pi q\bar q}/(8\pi) }
\def \be {\begin{equation}}
\def \ba {\begin{eqnarray}}
\def \ee {\end{equation}}
\def \ea {\end{eqnarray}}
\def\bea{\begin{eqnarray}}
\def\eea{\end{eqnarray}}
\def \gap {{\rm gap}}
\def \gapp {{\rm \overline{gap}}}
\def \gappp {{\rm \overline{\overline{gap}}}}
\def \im {{\rm Im}}
\def \re {{\rm Re}}
\def \Tr {{\rm Tr}}
\def \P {$0^{-+}$}
\def \S {$0^{++}$}
\def\zpp{$0^{++}$}
\def\fz{$f_0(980)$}
\def\az{$a_0(980)$}
\def\Kz{$K_0^*(1430)$}
\def\fzz{$f_0(1370)$}
\def\fzzz{$f_0(1200-1300)$}
\def\azz{$a_0(1450)$}
\def\ss{$ s\bar s $}
\def\uu{$u\bar u+d\bar d$}
\def\qq{$q\bar q$}
\def\KK{$K\bar K$}
\def\sig{$\sigma$}\def\kap{$\kappa$}
\def\lsim{\;\raise0.3ex\hbox{$<$\kern-0.75em\raise-1.1ex\hbox{$\sim$}}\;}
\def\gsim{\raise0.3ex\hbox{$>$\kern-0.75em\raise-1.1ex\hbox{$\sim$}}}
\def\one{\bf 1}

\date{today}

The mesons with vacuum quantum numbers are known to be crucial for
a full understanding of the symmetry breaking mechanisms in QCD,
and presumably also for confinement. The  lightest scalar mesons
have been controversial since their first observation over thirty
years ago. Due to the complications of the nonperturbative strong
interactions there is still no general agreement as to where are
the $q\bar q$ states, whether there is necessarily a glueball
among the light scalars, and whether some of the too numerous
scalars are multiquark, $K\bar K$ or other meson-meson bound
states.  The main problem is that there are too many\cite{kyoto}
light scalars below 1.5 GeV.

A  likely solution\cite{CloseNT} is that in addition  to a \qq\
nonet and a glueball above 1.2 GeV, there is another nonet of more
complicated nature  below 1 GeV ($\sigma(600),\ a_0(980)$,
$f_0(980)$, $\kappa(\sim 800)$\cite{kappa}, i.e., 18 scalar states
in all. The latter nonet should have large 4-quark and meson-meson
components. There is a heated current debate as to whether the
\sig\ and especially the \kap\ really are true resonances or just
due to very strong attractions in the $\pi\pi$ and $K\pi$
channels. Here we do not want to enter into this debate, we shall
only assume that one can approximately model these effects by
effective fields. We discussed this with  Close in in more detail
in a recent review\cite{CloseNT}.

The \sig\ is sometimes called a Higgs boson of strong interactions
since in a simple NJL model and in a linear sigma model the \sig\
acts like a Higgs giving the constituent $u$ and $d$ quarks most
of their mass, and one has the celebrated Nambu relation $m_\sigma
=2m_u^{const}$. But, in such models one generally breaks only a
global symmetry spontaneously. For a true analogy with the Higgs
mechanism one should have a local symmetry which is broken
spontaneously or dynamically. Can one construct\cite{nils0201171}
such a model?

I shall argue  that two coupled linear sigma models may provide a
first step for an understanding of this and of  such a
proliferation of 18 light scalar states. After gauging a hidden
$U(3)$ symmetry one can then look at the lightest scalars as
Higgs-like bosons for the nonperturbative low energy strong
interactions.

 Let me first remind the reader of the simple $U(N_f)\times
U(N_f)$ linear sigma model\cite{u3u3} which includes
 one scalar and one pseudoscalar multiplet. As well known this agrees with
 chiral perturbation theory at the lowest order in $p^2$\cite{schech},
 but includes explicit scalars.
 The scalar nonet is put
 into the hermitian part of a
$3\times 3$ matrix $\Phi$ and the pseudoscalar nonet into the
anti-hermitian part of $\Phi$. One has $\Phi=S+iP=
\sum_{a=0}^8(\sigma_a+ip_a)\lambda_a/\sqrt 2$, where $\lambda_a$
are the Gell-Mann matrices, and  $\lambda_0 = (2/N_f)^{1/2} {\bf
1}$.
 Then  the potential \be V(\Phi) =-\frac 1 2 \mu^2{\rm Tr}
[\Phi\Phi^\dagger] +\lambda {\rm Tr}[\Phi\Phi^\dagger
\Phi\Phi^\dagger] +\lambda' ({\rm Tr}[\Phi\Phi^\dagger])^2+{\cal
L_{SB}}, \ee where $\lambda '$ is a small parameter compared to
$\lambda$ (which breaks the scalar singlet mass from that of the
octet) and where ${\cal L_{SB}}$ contains a flavor symmetry
breaking term $\propto$Tr$(\Phi M_q+M_q\Phi^\dagger)$ (where $M_q$
is the diagonal matrix composed of $m_u,m_d,m_s$), and an $U_A(1)$
breaking term $\propto($det$\Phi$+det$\Phi^\dagger ) $, is not a
too bad representation of the lightest pseudoscalars and scalars,
already at the tree level. If five of the six parameters are fixed
by the experimental $m_\pi^2, m_K^2, (m_\eta^2+m_{\eta '}^2)$,
$f_\pi$ and $f_K$, one finds with a small sixth parameter
($\lambda '$) the scalar nonet to be near 1 GeV (a very broad
\sig\ near 650 MeV, an $a_0$ at 1040 MeV, an $f_0$ near 1200 MeV,
and a very broad \kap\ near 1120 MeV\cite{lsm}). This is quite
reasonable considering that unitarizing a similar model can, and
in fact and does\cite{NAT12}, shift these states in the second
sheet by hundreds of MeV. The essential features we recall here is
that neglecting the $U_A(1)$ term  one has, after a shift to the
minimum $\Phi\to\Phi +v\one$ (where $v^2=m^2/(4\lambda) + {\cal
O}(m_q)$, a nearly massless pseudoscalar nonet of squared mass of
${\cal O}(m_q)$ and a massive scalar nonet of squared mass
$=2m^2+{\cal O}(m_q)$.

Now, for two scalar nonets in a chiral model we need  two such
$3\times 3$ matrices $\Phi$ and $\hat \Phi$. (Let the scalar \qq\
states above 1 GeV be in $\Phi$, and let those below 1 GeV be in
$\hat\Phi$). Then model both  $\Phi$ and $\hat\Phi$ by a gauged
 linear sigma
model, but with different sets of parameters ($\mu^2, \lambda$ )
and ($\hat\mu^2, \hat\lambda$). For $\Phi$ without any symmetry
breaking nor a $\lambda'$ term we have simply \be
 {\cal L}(\Phi ) =  \frac 1 2 {\rm Tr}
[D_\mu\Phi D_\mu\Phi^\dagger] +\frac 1 2 \mu^2{\rm Tr} [\Phi
\Phi^\dagger] -\lambda {\rm Tr}[\Phi\Phi^\dagger
\Phi\Phi^\dagger], \ee and similarly for $\hat\Phi$: \be \hat{\cal
L}(\hat\Phi ) = \frac 1 2 {\rm Tr} [D_\mu\hat \Phi
D_\mu\hat\Phi^\dagger] +\frac 1 2 \hat\mu^2{\rm Tr} [\hat\Phi
\hat\Phi^\dagger] -\hat\lambda {\rm Tr}[\hat \Phi\hat\Phi^\dagger
\hat\Phi\hat \Phi^\dagger].\label{Lagtot}\ee
  Neglect to begin with the gauging.
  We have doubled the spectrum and initially we have two scalar,
and two pseudoscalar multiplets, altogether 36 states for three
flavors.

These lagrangians are invariant under a global  symmetry: $\Phi
\to L\Phi R $ and $\hat \Phi \to L\hat \Phi R $, where $L$ and $R$
are independent $U(3)=SU(3)\times U(1)$ transformations. If there
were no coupling between $\Phi$ and $\hat \Phi$ the symmetry would
be even larger as the $U(3)$ transformations on $\Phi$ could be
independent of those on $\hat \Phi$. We refer to that symmetry as
the {\it relative} symmetry. But, it is natural to introduce a
small coupling\cite{black}) between the two sets of multiplets,
which breaks this relative symmetry\cite{nils0201171}.

The full effective Lagrangian for both $\Phi$ and $\hat\Phi$ thus
becomes, \be {\cal L}_{tot}(\Phi ,\hat\Phi ) ={\cal L}(\Phi
)+\hat{\cal L} (\hat\Phi)+ \frac {\epsilon^2} 4 {\rm
Tr}[\Phi\hat\Phi^\dagger +h.c.].\ee If $\Phi_a$ is interpreted as
$q\bar q$ and $\hat \Phi_a$ as $q\bar qq\bar q$  states then the
$\epsilon^2$ term would allow for $q\bar q\to q\bar qq\bar q$
transitions\cite{bogli}.  This Lagrangian is still invariant under
the above $U(3)\times U(3)$ symmetry, but not under the relative
symmetry when $\Phi$ is transformed differently from $\hat\Phi$.

Now as a crucial assumption (differently from\cite{black}), let
both $\Phi$ and $\hat\Phi$ have vacuum expectation values (VEV),
such that $v=< \sigma_0>/\sqrt N_f \neq 0$ and $\hat v= <\hat
\sigma_0>/\sqrt N_f \neq 0$ even if $\epsilon=0$. Then one has
$v^2(\epsilon)=(\mu^2+\epsilon^2\hat v/v)/(4\lambda)$, and $\hat
v^2(\epsilon)=(\hat \mu^2+\epsilon^2 v/\hat v)/(4\hat\lambda)$. If
$\epsilon$ would vanish all pseudoscalars would be massless, but
with $\epsilon\neq 0$ the $2\times 2$ submatrix between two
pseudoscalars with same flavor becomes: \be m^2 (0^{-+})= {\left(
\begin{array}{cc}
4\lambda v^2(\epsilon)-\mu^2 & -\epsilon^2  \\
-\epsilon^2& 4\hat\lambda\hat v^2(\epsilon)-\hat\mu^2
 \\
        \end{array}
        \right )
        =+\epsilon^2 \left( \begin{array}{cc}
\hat v/v& -1          \\
-1         & v/\hat v \\
        \end{array}
\right )} ,\ee which is diagonalized by a rotation $\theta=\arctan
(v/\hat v)$, such that the eigenvalues are 0 and $\epsilon^2v\hat
v/(v^2+\hat v^2)$: \be \left(
\begin{array}{cc}
c & -s  \\
s& c \\
        \end{array}\right )m^2 (0^{-+})\left( \begin{array}{cc}
c & s  \\
-s& c \\
        \end{array}\right )=
        \epsilon^2\frac{v^2+\hat v^2}{v\hat v}\left( \begin{array}{cc}
1 & 0  \\
0 & 0  \\
        \end{array}\right ).      \ee
Here $s=\sin\theta\propto v$ and $c=\cos\theta\propto \hat v$.
Thus the two originally massless pseudoscalar nonets mix through
the $\epsilon^2$ term, with a mixing angle $\theta$, such that one
nonet remains massless, while the other nonet obtains a mass
$\epsilon^2(v^2+\hat v^2)/v\hat v $. This is, of course, just what
is expected, since we still have one exact overall $U(3)\times
U(3)$ symmetry, while the relative symmetry is broken through the
$\epsilon$ term.

The approximation is valid only if neither $v$ nor $\hat v$
vanishes. Thus one has one massive $|\pi>$ and one massless $|\hat
\pi
>$ would-be pseudoscalar multiplet. Denoting the the original
pseudoscalars $|p>$ and $|\hat p>$, we have $|\pi>=c|p>-s|\hat
p>$, and $|\hat \pi>=s|p>+c|\hat p>$. The mixing angle is
determined entirely by the two vacuum expectation values, and is
large if $v$ and $\hat v$ are of similar magnitudes, independently
of how small $\epsilon^2$ is, as long as it remains finite. On the
other hand the scalar masses and mixings are only very little
affected if $\epsilon^2/(\mu^2-\hat\mu^2)$ is small. They are
still close to $\sqrt 2 \mu $ and $\sqrt 2 \hat \mu$ as in the
uncoupled case.

In order that this should have anything to do with reality, one
must of course get rid of the massless Goldstones.
 By gauging the overall axial symmetry
($\Phi\to H\Phi H$ and $D_\mu=\partial_\mu-i
g/2(\lambda_aA_a+A_a\lambda_a)$) and reparameterizing the fields
%\be\Phi= v{\bf 1}+ (\sigma_a+ip_a)\frac {\lambda_a}{\sqrt 2} \to
%\exp[\frac {is\hat \pi_a\lambda_a} {v\sqrt 2}][ v{\bf
%1}+(\sigma_a+ic\pi_a)\frac {\lambda_a}{\sqrt 2}], \ee \be\hat
%\Phi=\hat v{\bf 1} +(\hat \sigma_a+i\hat p_a)\frac
%{\lambda_a}{\sqrt 2}\to \exp[\frac {ic\hat \pi_a\lambda_a} {\hat
%v\sqrt 2}][\hat v{\bf 1}+(\hat
%\sigma_a-is\pi_a)\frac{\lambda_a}{\sqrt 2}]. \ee
\be\Phi= v{\bf 1}+ (\sigma_a+ip_a)\frac {\lambda_a}{\sqrt 2} \to
H'[ v{\bf 1}+(\sigma_a+ic\pi_a)\frac {\lambda_a}{\sqrt 2}]H', \ee

\be\hat \Phi=\hat v{\bf 1} +(\hat \sigma_a+i\hat p_a)\frac
{\lambda_a}{\sqrt 2}\to H'[\hat v{\bf 1}+(\hat
\sigma_a-is\pi_a)\frac{\lambda_a}{\sqrt 2}]H'. \ee Here $H'$ is a
fixed gauge for the axial symmetry
 \be
 H'=
 \exp[\frac {i\hat \pi_a\lambda_a} {2\sqrt
{2(v^2+\hat v^2)}}])  .\ee
 The validity of these reparametrizations can be  seen most easily
 by  expanding  $H'=$
 $1+i\frac{s}{2v} \frac{\hat \pi_a\lambda_a}{\sqrt 2} ... $
 $ =1+i\frac{c}{2\hat v} \frac{\hat \pi_a \lambda_a}{\sqrt 2} ... $
 Thus by choosing a special gauge for the hidden symmetry $H$ the
$\hat \pi_a$ fields vanish from the spectrum. The axial symmetry
$H$ remains as a hidden symmetry while the $\hat \pi$ fields are
gauged away. But, these degrees of freedom enter instead as
longitudinal axial vector mesons and give these  mesons (an extra)
mass ($m_A^2=2g^2(v^2+\hat v^2)$.
 This is  like the conventional Higgs mechanism and it
 has  similarities to the original Yang-Mills theory and
the work of Bando et al.\cite{bando} on hidden local symmetries,
in that  mesons are gauge bosons, but is  different both in the
scalar particle spectrum and in the realization of the hidden
symmetry. The axial vector-pseudoscalar-scalar couplings ($APS$)
can be read off from the lagrangian. For the $\sigma$ multiplet
one finds
 $gcA_{\mu,a}[\pi_b\partial_\mu
\sigma_c+\sigma_b\partial \pi_c]{\rm
Tr}(\lambda_a\lambda_b\lambda_c)_+/4$, while for the $\hat \sigma$
multiplet  $c$ is replaced by $s$. Other trilinear couplings also
follow, in particular for scalar to 2 pseudoscalar couplings
($SPP$) one has: $g_{\hat \sigma_a \pi_b \pi_c}= v cs {\rm
Tr}(\lambda_a\lambda_b\lambda_c)_+/\sqrt 2$ and, $g_{ \hat
\sigma_a \pi_b \pi_c}/g_{ \sigma_a \pi_b \pi_c} = v/ \hat v=\tan
\theta$.

 Now, having gauged away the
massless Goldstones one can interpret the massive pseudoscalars as
the physical pseudoscalars. The would-be axial current related to
the overall hidden symmetry is like the $\hat \pi $ gauged away,
while the explicitly broken relative symmetry defines a current
which is only "partially conserved" when $\epsilon$ differs from
0. Denoting  the axial vector current obtained from ${\cal L}(\Phi
) $ by $j_{A\mu,a}=\sqrt N_f v\partial_\mu p_a + ...$ and the one
from $\hat{\cal  L}(\hat \Phi )$ by $ \hat j_{A\mu}=\sqrt N_f\hat
v\partial_\mu \hat p_a + ...$, then both currents would before
gauging be conserved if $\epsilon=0$, because of the masslessness
of both $0^-$ nonets. Adding the $\epsilon$ term the sum
$j_{A\mu}+\hat j_{A\mu}$ would still be exactly conserved, because
of the $H$ symmetry and since it would be $\propto
\partial_\mu \hat \pi$, but this current is like the $\hat \pi$ gauged away.
On the other hand $j_{A\mu,a}$ or $\hat j_{A\mu,a}$ alone is only
"partially conserved", $\partial_\mu j_{A\mu,a}$= $-\partial_\mu
\hat j_{A\mu,a}=\sqrt N_f v\hat v/\sqrt(v^2+\hat v^2)m^2_\pi
\pi_a$, because  the $\epsilon^2$ term explicitly breaks the
relative symmetry when the $\pi$ nonet obtains mass.  Identifying
this with PCAC one has \ba
f_\pi = \sqrt N_f v\hat v/\sqrt (v^2+\hat v^2) , \\
m_\pi^2=\epsilon^2(v^2+\hat v^2)/v\hat v  . \ea Comparing this
with the conventional relation $m_\pi^2=2B\hat m_q$, where $\hat
m_q$ is the average chiral quark mass one sees that  $\epsilon^2$
should be proportional to  $\hat m_q$. In fact a natural way to
break flavor symmetry is obtained by replacing \be \frac
{\epsilon^2} 4 {\rm Tr}[\Phi\hat\Phi^\dagger +h.c.]\to
\frac{B'}{2}{\rm Tr}[\Phi M_q \hat\Phi^\dagger +h.c.],
\label{sbt}\ee

 Then
 the $v{\bf 1}$ and $\hat v{\bf 1}$ will be replaced by a
diagonal matrix with elements $v_{i\bar i}$ which includes
 corrections due to unequal quark masses and satisfy
 $v_{i\bar i}^2=(\mu^2+2B'm_{q_i}\hat v_{i\bar i}/v_{i\bar i})/(4\lambda)$
 $i\bar i= u\bar u,d\bar d,s\bar s$ and a similar equation for $\hat v_{i\bar i}^2$.
 For small $m_{q_i}$ one then recovers the usual relations that
squared pseudoscalar masses are $\propto (m_{q_i} +m_{q_j})$,
whereas the two scalar nonets as well as the vectors get split by
the equal spacing rule.

If the $\hat \sigma$ nonet is predominantly of the 4-quark form of
Jaffe\cite{Ja77} it should, with the appropriate symmetry breaking
term, before unitarization obey the inverted mass spectrum  with
the $a_0(980)$ and $f_0(980)$ as the heaviest followed by the
$\kappa$ and the $\sigma(600)$.

Also, if $\hat v_{i\bar i} << v_{i\bar i}$ one recovers for the
lighter multiplet ($\hat \sigma_a$) the predictions of the simple
$U(3)\times U(3) $ model discussed above in connection with Eq.
(1). (The term $\det \Phi + h.c$  in ${\cal L}_{SB}$ of Eq. (1)
would  here be replaced by $\propto \det\Phi\hat\Phi + h.c$.) From
the fact that $\hat \mu<\mu$ and that the $SPP$ couplings of the
lower multiplet should be larger than the heavier one expects, in
fact, that $\hat v_{i\bar i} < v_{i\bar i}$ or $\tan\theta_i
>1$, but it is crucial that both $v_{i\bar i}\neq 0 $, and $\hat v_{i\bar i}\neq 0$.

 The main prediction of this scheme
is that one have doubled the light scalar meson spectrum, as seems
to be experimentally the case. Of course in order to make any
detailed comparison with experiment one must include loops and
unitarize the model, which is not a simple matter as the couplings
are very large.

The dichotomic role of the pions in conventional models, as being
at the same time both the Goldstone bosons and the \qq\
pseudoscalars, is here resolved in a particularly simple way: One
has originally two Goldstone-like pions, out of which only one
remains in the spectrum, and which is a particular linear
combination of the two original pseudoscalar fields.

Both of the two scalar multiplets remain as physical states
 and one of these (formed by the $\sigma(600)$ and the $ a_0(980)$ in the case of  two flavors),
 or the $\sigma,\ a_0(980),\ f_0(980)$ and the $ \kappa\  $ in the case of three flavors
can then be looked upon as effectively a Higgs multiplet of strong
nonperturbative interactions when a hidden local  symmetry is
spontaneously broken.

 One may ask is there any other source for the symmetry
breaking term (12), except for the chiral quark masses put in by
hand? The Syracuse group\cite{black} argues for instanton effects.
 Another way of reasoning is that with quarks and quark loops
there would be anomalous couplings $AVV$ for each
flavor\cite{wein}. Anomaly related loops (like $P\to VV\to P$)
could then be another source of the symmetry breaking.

\section{Acknowledgements}
 Support  from EU-TMR program, contract
CT98-0169 is gratefully acknowledged. I thank J. Schechter,
Syracuse for emphasizing that it should be the axial vectors which
get mass through this mechanism.

\end{document}